\documentclass[preprintnumbers,amsmath,amssymb,floatfix,11pt,onecolumn,prd,superscriptaddress,nofootinbib,showpacs,showkeys]{revtex4}
\usepackage{graphicx}
\usepackage{epsfig}
\usepackage{bm}
\usepackage{amsfonts}

\def\ben{\begin{equation}}
\def\een{\end{equation}}

 \def\bd{\begin{document}} \def\ed{\end{document}}
\def\ds{\documentstyle} \let\fr=\frac \let\bl=\bigl \let\br=\bigr
\let\Br=\Bigr \let\Bl=\Bigl
\let\bm=\bibitem
\let\na=\nabla
\let\pa=\partial \let\ov=\overline
\newcommand{\be}{\begin{equation}}
\newcommand{\ee}{\end{equation}}
\def\ba{\begin{array}}
\def\ea{\end{array}}
\def\ft#1#2{{\textstyle{\frac{\scriptstyle #1}{\scriptstyle #2} } }}
\def\fft#1#2{{\frac{#1}{#2}}}
\def\del{\partial}
\def\vp{\varphi}
\def\sst#1{{\scriptscriptstyle #1}}
\def\oneone{\rlap 1\mkern4mu{\rm l}}
\def\td{\tilde}
\def\wtd{\widetilde}
\def\ie{{\it i.e.\ }}
\def\dalemb#1#2{{\vbox{\hrule height .#2pt
        \hbox{\vrule width.#2pt height#1pt \kern#1pt
                \vrule width.#2pt}
        \hrule height.#2pt}}}
\def\square{\mathord{\dalemb{6.8}{7}\hbox{\hskip1pt}}}
\newcommand{\ho}[1]{$\, ^{#1}$}
\newcommand{\hoch}[1]{$\, ^{#1}$}
\newcommand{\bea}{\setlength\arraycolsep{2pt} \begin{eqnarray}}
\newcommand{\eea}{\end{eqnarray}}
\newcommand{\ra}{\rightarrow}
\newcommand{\lra}{\longrightarrow}
\newcommand{\Lra}{\Leftrightarrow}
\newcommand{\bp}{\tilde \beta^\prime}
\newcommand{\tr}{{\rm tr} }
\newcommand{\Tr}{{\rm Tr} }
\def\0{{\sst{(0)}}}
\def\1{{\sst{(1)}}}
\def\2{{\sst{(2)}}}
\def\3{{\sst{(3)}}}
\def\4{{\sst{(4)}}}
\def\5{{\sst{(5)}}}
\def\6{{\sst{(6)}}}
\def\7{{\sst{(7)}}}
\def\8{{\sst{(8)}}}
\def\m{{\sst{(m)}}}
\def\n{{\sst{(n)}}}
\def\cA{{{\cal A}}}
\def\cB{{{\cal B}}}
\def\cF{{{\cal F}}}
\def\cG{{{\cal G}}}
\def\cH{{{\cal H}}}
\def\tV{\widetilde V}
\def\tW{\widetilde W}
\def\tH{\widetilde H}
\def\tE{\widetilde E}
\def\tF{\widetilde F}
\def\tA{\widetilde A}
\def\im{{{\rm i}}}
\def\tY{{{\wtd Y}}}
\def\ep{{\epsilon}}
\def\vep{{\varepsilon}}
\def\bD{{{\bar D}}}
\def\R{{{\mathbb R}}}
\def\C{{{\mathbb C}}}
\def\H{{{\mathbb H}}}
\def\CP{{{\mathbb C}{\mathbb P}}}
\def\RP{{{\mathbb R}{\mathbb P}}}
\def\Z{{{\mathbb Z}}}
\def\bA{{{\mathbb A}}}
\def\bB{{{\mathbb B}}}
\def\bC{{{\mathbb C}}}
\def\bD{{{\mathbb D}}}
\def\bE{{{\mathbb E}}}
\def\bZ{{{\mathbb Z}}}
\def\Re{{{\frak{Re}}}}
\def\Im{{{\frak{Im}}}}
\def\cosec{{\,\hbox{cosec}\,}}
\def\Gm{{\Gamma_{\!\! -}}}
\def\Gp{{\Gamma_{\!\! +}}}
\def\stan{{standard }}
\def\nonstan{{supernumerary }}
\def\p{{\partial}}
\def\kdel#1{{\fft{\del}{\del#1}}}

\def\bog{{Bogomolny }}
\def\om{{\omega}}

\newcommand{\nnr}{\nonumber \\}
\newcommand{\pd}{\partial}
\newcommand{\ud}{\textrm{d}}
\newcommand{\dTH}{T^{\prime \, 0}_\textrm{H}}
\newcommand{\dOi}{\Omega^{\prime \, 0}_i}
\newcommand{\bx}{{\bf x}}
\begin{document}
\title{More on Superconductors via Gauge/Gravity Duality with Nonlinear Maxwell Field}

\author{\textbf{Davood Momeni}}
 \affiliation{Eurasian International Center for Theoretical Physics, Eurasian National University, Astana 010008, Kazakhstan.}
\author{\textbf{Muhammad Raza}}
\affiliation{Department of Mathematics, COMSATS Institute of
Information Technology (CIIT), Sahiwal campus, Pakistan}
\author{\textbf{Ratbay Myrzakulov}}
\affiliation{Eurasian International Center
for Theoretical Physics, Eurasian National University, Astana
010008, Kazakhstan.}
\begin{abstract}
 We have developed the recent investigations on the second-order phase transition in the holographic superconductor using the
probe limit for a nonlinearMaxwell field strength coupled to amassless scalar field. By analytical methods, based on the variational
Sturm-Liouville minimization technique, we study the effects of the spacetime dimension and the nonlinearity parameter on
the critical temperature and the scalar condensation of the dual operators on the boundary. Further, as a motivated result, we
analytically deduce theDCconductivity in the lowand zero temperatures regime. Especially in the zero temperature limit and in two
dimensional toy model, we thoroughly compute the conductivity analytically. Our work clarifies more features of the holographic
superconductors both in different space dimensions and on the effect of the nonlinearity in Maxwell’s strength field.
\end{abstract}
 \keywords{ High-$T_C$ superconductors theory; Gauge/Gravity duality.}
 \maketitle

\section{Introduction}
In the recent years, using the holographic picture of the
world, the AdS/CFT (anti de Sitter/conformal field theory)
correspondence [1-3] has been applied to study some
strongly correlated systems in condensed matter physics,
especially for strongly coupled systems with the scaleinvariance.
Particularly, people studied the low temperature,
quantum critical systems near critical point (see, e.g., [4,
5] and references therein). The critical phenomena, which
happen here, is a second-order phase transition from normal
phase to the superconducting phase, in which below
a specific temperature 𝑇
𝑐, the DC conductivity becomes
infinite. Such second-order phase transitions happen in the
high-temperature superconductors and can be described very
well by the AdS/CFT dictionary [6, 7]. From the classical
and phenomenological point of view, superconductivity,
in the high-temperature type II superconductors,formulated
using a phenomenological based Landau-Ginzburg Lagrangian. This
Lagrangian contains a general complex value scalar field Ψ,
plays the role of a condensate in a superconductive phase.
Basically, to have a scalar condensation in the boundary
quantumfield theory using CFT on the boundary of the bulk,Hartnoll et al [8] proposed a Lagrangian of an abelian gauge field 𝐴
𝜇and a complex scalar field with mass above the Breitenlohner-Freedman (BF) bound [9]. Later Gubser [10] stdied the hairy black holes and he showed that how the dual operators has the same temperature dependence as the condensation in superconductors.\par
The full description of the superconductivity in the probe
limit or away this limit needs to provide the numerical
solutions of a couple of nonlinear differential equations. By
simplicity, they can be solved using the shooting approach
by expanding in series the functions and matching these by
varying the free parameters of the series in a typical point
between the horizon and the spatial infinity. Parallel to the
numerical studies, recently some analytical approaches have
been proposed to find the universal properties of secondorder
phase transitions in holographic superconductors [11–
18]. In particular, the authors in [18] used the variational
functional method. In
[18, 19] this analytical method has been used to 
calculate critical properties like temperature and critical exponent. The eigenvalue of this variational problem is a
function of the critical chemical potential 𝜇
𝑐$\mu_c$, and consequently,
it is related to the 𝑇
𝑐$T_c$. another types of super
criticality have been studied depending on the dual operators on boundary [20].
Also, one can apply this method to superconductors with
external magnetic fields[21, 22].
Furthermore, a number of aspects of external and bulk
magnetic fields in holographic superconductors have been
investigated [23-26]. The phase transition can be interpreted
in terms of the string interactions [27-31]. The effects of the
nonlinear electrodynamics in the holographic superconductors
have been investigated recently [32-36].
There are many interests in the modified gravity theories.
For example, on Gauss-Bonnet and Weyl corrected
superconductors, on which, we are working with a higher
derivative corrected bulk black hole, like Weyl corrections
[37] numerically. Furthermore, we have studied the Weyl
corrections to the superconductors analytically [38, 39].
Moreover, we showed that there exists a family of p-wave
holographic withWeyl corrections [40].
In the present paper, we would like to study the $((𝑑d-
2) + 1)$-dimensional holographic superconductor in the
probe limit for a power law Maxwell field strength $(F^{\mu\nu}F_{\mu\nu})^{\delta}$
coupled to a scalar field. We focus just on the s-wave cases.
We must clarify the motivation of the s-wave approximation
in holographic models of superconductors. In the relativistic
models of the gravity, it is highly known that s-wave approximation
is not a good approximation, for example, in the
cosmological models and black holes [41]. The meaning of
the s-wave here does not back to the reduction of the action
from four to two dimensional like the dilatonic action from
the four dimensional spinor (Majorana) action. We mean
by s-wave, in the context of holographic superconductors, a
scalar order parameter, whose expectation value breaks the
U(1) but not rotational symmetry.Moreover, we can have the
Yang-Mills fields with 𝑆𝑈(2) symmetry which additionally
they can generate another symmetry breaking of an axial
vector type. The last case resembles the p-wave models. We
mention here that the three dimensional non-linear model of
the superconductors,which we used in this paper, is a realistic
model and it will be more interesting that we can find a direct
relation between this nonlinear model and the results of a
higher dimensional model, by a principle like the detailed
balance.
Another additional point is to restrict ourselves just to the
case of a single horizon. The problem of the multihorizon
cases needs more investigation, for example, the case of
the Nariai black holes. The theory here will be so different.
This later appeared in the lower dimensional models. For
example, the case of the quantum corrected BTZ like black
hole is a good example [42]. In the holographic set up
for superconductors one must identify a temperature in
his gravitational bulk model to the CFT temperature on
the boundary. If the black hole has only one horizon, in
this case, we can use the Hawking-Bekenstein (horizon)
or Kodama- Hayward temperature [43] as a reasonable
candidate. But if our asymptotically AdS bulk has more
than one horizon, for example, in the case of the charged
BTZ like black holes, then we take the temperature of the
real physical horizon (the temperature which is obtained by
calculation the surface gravity of the biggest null hypersurface
orthogonal surface) as the candidate for temperature of the
CFT. In fact the effects of the quantum corrections and
chargedMaxwell field on the background of the bulk are very
interesting problems and can be investigated in more details.
Also it is possible to relate the instability of such charged
dilaton configurations in the AdS spacetime [44] to the
symmetry breaking mechanism of the superconductors.The
idea has motivation enough as a new work. In this paper we
investigate analytically the effect of the spacetime dimension
𝑑 and the power $\delta$ on the critical temperature 𝑇
𝑐. Although
our problem is the especial massless case of the model
which has been investigated recently [45],\cite{49}, we study these
corrections to the superconductors analytically. Additionally,
we want to compute the DC conductivity for this kind of
the superconductor using the perturbation method. In this
approach, we apply an external linear electromagnetic field.
This field is periodic in time. By calculating, the response
in the first order linear approximation, we compute the
conductivity for the low temperature case, especially for zero
temperature configuration.
Our plan in this paper is as the following. In Section 2, we
clarify our motivation for considering the nonlinearMaxwell
action instead of the linear theory. In Section 3, introduce
ourmodel for holographic superconductors. In Section 4, we
apply the variational method to obtain the critical temperature
of the system. In Section 5, we calculate the critical
exponent for the condensation operator. In Sections 6 and 7
we compute the conductivity for low and zero temperature
cases.We summarize and conclude in the final section.
\section{Motivation for nonlinear Maxwell effects in holographic superconductors}

Thelinear approximations in themathematical physics, as we
know, have limitations, both in predictions of the model and
especially on matching with the full description of the model
using the numerical results.The linearMaxwell theory fails in
some domains, and it is needed to consider the general form
of the Lagrangian instead of the linear one. The Lagrangian
of the Maxwell model is:
\begin{eqnarray}
\mathcal{L}_{M}=-\frac{1}{4} F^2\label{maxwell},
\end{eqnarray}
where $F^2=F^{\mu\nu}F_{\mu\nu}$.
 A natural extension of (1) is obtained by
replacing a general function of 𝐹 in the form $\Phi(F)$. Recently, it
has been shown that such nonlinear general forms have a rich
family of black holes in $f(R)$ gravity [46]. There are different
reasons for investigating these forms. The oldest one may be
the Born-Infeld (BI) alternative for linear Maxwell’s theory.In string theory language, this BI action can be replaced by
the tachyonic action. The BI Lagrangian reads [47]:
\begin{eqnarray}
\mathcal{L}_{BI}=\eta^2 (1-\sqrt{1+\frac{F^2}{2\eta^2}})\label{bi}.
\end{eqnarray}
Here $\eta$ is the string's tension parameter. This form reduces to
the (\ref{maxwell}) in the limit of $\eta\rightarrow\infty$. Indeed, we can expand in series
(2) in the following form:
\begin{eqnarray}
\mathcal{L}_{BI}= \sum_{n=1}^{\infty}c_n \eta^{2(1-n)}F^{2n}.
\end{eqnarray}
The leading order term $n=1$ is just in form of (\ref{maxwell}).
Even, if we don't work with BI theory, this nonlinearity meets us
from a geometrical point of view. Suppose that we want to write a
conformal invariance (CI) Lagrangian, constructed from the $U(1)$
gauge fields $A^{\mu}$. Such CI is the invariance of the whole
theory under geometrical transformation $g_{\mu\nu}\rightarrow
e^{2\sigma} g_{\mu\nu}$ in a $d\geq4$ Riemannian manifold without
torsion or non metricity fields. However, previously, a version of
such Lagrangian has been found in Weitznbock  spacetime with torsion
and non-metricity fields\cite{torsion}. It is easy to show that the
Lagrangian $\mathcal{L}\propto F^{d/2}$ is invariant under CI
transformations. When $d=4$, the proper Lagrangian is
$\mathcal{L}\propto F^{2}$ but in $d=5$ the suitable form is
$\mathcal{L}\propto F^{5/2}$. The last form belongs to the
nonlinear, non integer Maxwell family. Thus from geometrical view,
the nonlinearity is welcome in our Lagrangian dynamical theory.
Further, even if we don't know any on BI or CI, when we are working
with vacuum effects , there is a simple generalization of Maxwell
Lagrangian in a logarithmic form
\begin{eqnarray}
\mathcal{L}_{log}=-\eta^2 \log(1+\frac{F^2}{4\eta^2}).
\end{eqnarray}
As BI case, in the limit of $|F|<<\eta$, by expanding this equation
 the leading order term $n=1$, is the linear
theory(\ref{maxwell}). In brief, according to the above discussion,
it seems that, consideration of the nonlinear effects of the Maxwell
field as $F^\delta$ are important, and the significant differences
between the usual linear theory $F^2$ and nonlinear theory
$F^\delta$ can be shown. Holographic
superconductors provide a rich background for testing such
type of new physics. In this paper, we will describe a $d-1$
dimensional holographic superconductor (HSC) via 𝑑-
dimensional gravity dual, described by a 𝑑$d$ dimensional AdS
black hole on the static patch.We set the Stuckelberg field to
zero and work with massless scalar fields.
\section{Field equations}

We write the following action [32-36] for a powerMaxwell
field , which it is coupled minimally to a massless scaler
field 𝜓 in a 𝑑d-dimensional asymptotic $AdS_d$𝑑 spacetime [45]
\begin{eqnarray}
S=\int d^d x
\sqrt{-g}[R+(d-1)(d-2)-\xi(F^{\mu\nu}F_{\mu\nu})^\delta-|D_\mu \psi|^2)\label{action}
\end{eqnarray}
Here, we set the radius of the AdS, L=1,
$D_{\mu}=\partial_{\mu}-ieA_{\mu}$,
$F_{\mu\nu}\equiv2\partial_{[\mu}A_{\nu]}=\partial_{\mu}A_{\nu}-\partial_{\nu}A_{\mu}$,
$A_{\mu}$ is  $U(1)$ vector potential. The exponent (power) $\delta$
can be non integer as we explained it in the previous section.
Further, when we work with Maxwell linear electrodynamics
$\delta=1,d=4$, then $\xi=\frac{1}{4}$. However, in our case with
$\delta\neq1$, in general, $\xi\in\mathcal{R}$ remains as a free
parameter in our model. We set the electric charge $e=1$. In normal
phase and in the absence of the scalar field we put $\psi=0$. This action has been used before in literature
for full description of the nonlinear Maxwell field in the
gravitational action. In the probe limitwhen thematter action
and the gravitational part decouple, the system of the field
equations has an exact solution which will be discussed here.
The solution of the generalized Maxwell-Einstein equations
is the simple 𝑑-dimensional static Reissner-Nordstrom-Anti
de-Sitter (RNAdS) black hole with the following metric
form:
\begin{eqnarray}
g_{\mu\nu}=diag(-f(r),\frac{1}{f(r)},r^2\Sigma_{d-2})\label{metric},
\end{eqnarray}
where $\Sigma_{d-2}$ is the metric on a $(d-2)$ dimensional sphere and the metric function is,
$$f(r)=r^2(1-(\frac{r_{+}}{r})^{d-1})$$
 Here  $r_+$ is the black hole
horizon which in general is the largest root of the algebraic
equation $f(r_{+})=0$. If we set $\delta=1$ in (\ref{action}), we
recover the usual  holographic superconductors. In the probe limit
by ignoring the backreaction effects of the matter fields in the
background metric $g_{\mu\nu}$, and with spherically symmetric
static metric and by choosing a suitable gauge fixing for the gauge
field $A_{\mu}$ we can take the functions
$(\psi,\phi)\in\mathcal{R}$  and one variable functions. We assume
that the functions $\psi$ and $\phi$ have finite numbers of poles on
the real axis. It means the analytical solutions are
the forms of the Gauss-hypergeometric functions. To obtain the field
equations, we assume that $A_{\mu}=\phi(r)\delta_{\mu t}$. So, the
non zero components of the $F_{\mu\nu}$ reads as
$$
F_{rt}=-F_{tr}=\phi'.
$$
Hence, we have
$$
F^{\mu\nu}F_{\mu\nu}=-2\phi'^2.
$$
Also, the Ricci $R$ of a d-dimensional spacetime with spherical symmetry reads
$$
R=-\Big(f''+\frac{2(d-2)f'}{r}-\frac{(d-2)(d-3)f}{r^2}\Big).
$$
Further, we compute
$$
|D_\mu \psi|^2=f\psi'^2-\frac{\phi'^2}{f}.
$$
Finally, by plugging the above expressions into the action
(\ref{action}), making a partial integration, we get the following
effective Lagrangian
\begin{eqnarray}
\mathcal{L}=r^{d-2}\Big[-\Big(f''+\frac{2(d-2)f'}{r}-\frac{(d-2)(d-3)f}{r^2}\Big)+(d-1)(d-2)-\xi(-2\phi'^2)^\delta-\Big(f\psi'^2-\frac{\phi'^2}{f}\Big)\Big]\label{L}
\end{eqnarray}
As a first step, it is necessary to eliminate the $f''$ term, by
integration part by part. After it, to write the field equation, we
use the Euler-Lagrange equation as the following
$$
\frac{d}{dr}(\frac{\partial \mathcal{L}}{\partial
q_{,r}})=\frac{\partial \mathcal{L}}{\partial q},\ \
q=\{\phi,\psi\}.
$$
 The field equations,
derived from the (\ref{L}) reads
\begin{eqnarray}
\psi''+(\frac{f'}{f}+\frac{d-2}{r})\psi'+\frac{\phi^2}{f^2}\psi=0\label{feq1}\\
 \phi''+\frac{d-2}{2\delta-1}\frac{\phi'}{r}-C_\delta\frac{\psi^2}{\xi
 f}\phi
(\phi')^{2(1-\delta)}=0\label{feq2},
\end{eqnarray}
here $C_\delta=((-2)^{2-\delta}\delta(2\delta-1))^{-1}$. To avoid
the pure complex numbers in our field equations, we assume that
$\delta\neq 2-\frac{1}{2N}$. The case with $\delta=1, D=4$ is the
usual  four-dimensional HSC describes the
three dimensional superconductors.\\
These field equations, are the special massless case of the model
which has been investigated recently
\cite{powermaxwell,powermaxwell2,49}. To avoid from the diverging near
the singularity $f(r_{+})=0$ we write the boundary conditions for
(\ref{feq1}),(\ref{feq2}) by
$$
\phi(r_{+})=\psi'(r_{+})=0.
$$

 The asymptotic solutions for system
(\ref{feq1}, \ref{feq2}) on the AdS boundary $r\rightarrow\infty$,
are
\begin{eqnarray}
&&\psi=D_{-}+\frac{D_{+}}{r^{d-1}}\\
&&\phi=A-\frac{B}{r^\eta}\label{phiAB}
 \label{sol}
\end{eqnarray}
in (\ref{sol}),
$D_{\pm}=<\mathcal{O}_{\pm}>,\eta=\frac{d-2\delta-1}{2\delta-1}$, where
$<\mathcal{O}_{\pm}>$ is the expectation value of the CFT operator
$\mathcal{O}_{\pm}$ on the boundary, and the chemical potential and charge
density of the dual theory are $A=\mu,B=\rho^{(2\delta-1)^{-1}}$ respectively.\\
 We must clarify that why the form of the electric potential $\phi$ modified. In the asymptotic regime, we know that the metric function $f$ behaves like $f\sim r^2$. Also, the scalar field has the following asymptotic form $\psi\sim0$, so the (\ref{feq2}) gives us
\begin{eqnarray}
&&\phi''_{\infty}\sim-\frac{d-2}{2\delta-1}\frac{\phi_{\infty}'}{r}\label{phiinfty},
\end{eqnarray}
The solution (\ref{phiinfty}) reads
\begin{eqnarray}
&&\phi_{\infty}(r)=c_1+\frac{c_0(2\delta-1)}{2\delta-d+1}\frac{1}{r^{\frac{d-2\delta-1}{2\delta-1}}}.
\end{eqnarray}
This solution coincides completely on the solution presented in
(\ref{phiAB}). We mention here that the above function
$\phi_{\infty}(r)$ in the limit of the linear electrodynamic theory
$\delta=1$ has the true asymptotic form of $\phi_{\infty}(r)\sim
r^{3-d},d\neq2$. For $d=2$ the expression of $\phi_{\infty}(r)$ is
in the form of a diverging log term $\phi_{\infty}(r)\sim\log(r)$
and the application of the AdS/CFT fails. At least, we don't know
the unique and true dictionary of the AdS/CFT in this lower
dimensional bulk theory.

 Just remain to identify the parameters with the physical quantities in the dual theory, i.e. the chemical potential $\mu$ and the charge density $\rho$. We did it before. So the asymptotic behaviors have the same forms.\\
The asymptotic solutions for $\phi,\psi$ are the same as the previous expressions which it has been presented in the \cite{powermaxwell}. For
normalization purposes, we set $D_{-}=0$.

\section{Variational method}
To solve the solutions of the field equations given by
(\ref{feq1},\ref{feq2}) the well-known technique is solving them by
numerical algorithms. However, from these numerical solutions, it is
not so easy and straightforward to read $<\mathcal{O}>$. Another
method is using the matching method. It is a potentially powerful
method. Even so, the results must be interpreted very carefully near
the boundaries. The first step for solving
system(\ref{feq1},\ref{feq2}) using variational approach
\cite{Siopsis}, is rewriting the equations in a new dimensionless
coordinate $z=\frac{r_{+}}{r}$ in the following forms
\begin{eqnarray}
\psi''+(\frac{f'}{f}-\frac{d-4}{z})\psi'+(\frac{r_{+}}{z^2})^2\frac{\phi^2}{f^2}\psi=0\\
\phi''-\frac{\eta-1}{z}\phi'-C_\delta\frac{r_{+}^{2\delta}}{\xi
z^{4\delta}}\frac{\psi^2}{f}\phi (\phi')^{2(1-\delta)}=0.
\label{feqsz}
\end{eqnarray}
Now, $'=\partial_z$. Near the
critical point $T=T_c$, the following solution is valid
\begin{eqnarray}
\phi\approx\phi_B (1-z^\eta)
 \label{phi}
\end{eqnarray}
Here $\phi_{B}=\mu_c=\frac{\rho^{(2\delta-1)^{-1}}}{(r_{+})^{\eta+1}}$ is the value of the
$\phi$ at the horizon $r=r_{+}$ , $\delta\neq\frac{d-1}{2}$ and the critical point corresponds by the critical chemical potential $\mu=\mu_c$.
Further, we can write
\begin{eqnarray}
\psi\approx <O_{+}> (\frac{z}{r_{+}})^{d-1}F(z)
 \label{psi}
\end{eqnarray}
 near the AdS boundary $z\rightarrow0$ with $F(0)=1$, $F'(0)=0$. The
 function $F(z)$ satisfies the following second order Sturm-Liouville
 differential equation,

\begin{eqnarray}
&&[\mu(z)F'(z)]'-Q(z)F(z)+(\frac{\phi_B}{r_{+}})^2P(z)F(z)=0
 \label{feq}
\end{eqnarray}
where
\begin{eqnarray}
&&\mu(z)=z^{2d-2}(z^{d-1}-1)\\
&&P(z)=\mu(z)(\frac{1-z^\eta}{1-z^{d-1}})^2\\
&&Q(z)=-\mu(z)\Big[\frac{(d-1)(d-2)}{z^2}-\frac{d-1}{z}(\frac{2+(d-3)z^{d-1}}{z(1-z^{d-1})}+\frac{d-4}{z})\Big]
\end{eqnarray}
our strategy is to obtaining the minimum value of
$\frac{\phi_B}{r_{+}}$ from the minimization of the following
functional,
\begin{eqnarray}
\Theta(\delta,d)\equiv(\frac{\phi_B}{r_{+}})^2_{Min}=\frac{\int_{0}^{1}(\mu(z)F'(z)^2+Q(z)F^2(z))dz}{\int_{0}^{1}
P(z)F^2(z) dz}
 \label{functional}
\end{eqnarray}
The minimum of the critical temperature $T_c$ is obtained from the
$T=\frac{(d-1)r_{+}}{4\pi}$ . It reads
\begin{eqnarray}
T_c=\gamma \rho^{\eta+1}
 \label{tc}
\end{eqnarray}
and
$\gamma=\frac{d-1}{4\pi}((\frac{\phi_B}{r_{+}})_{Min})^{-(\eta+1)}$.
We use from the trial function $F(z)=1-\alpha z^2$ in
(\ref{functional}). It is useful to set $\rho=1$. The values of the
critical temperature $T_c$ for $d=4,5$ , by minimizing the functional
(\ref{functional}) with trial function F(z)  and using (\ref{tc})(in
case $\rho=1$)   are given by:

\noindent
For $d=4$, $\delta=1$: $T_c=0.0844$, for $\delta=3/4$, $T_c=0.1692$.

\noindent
For $d=5$, $\delta=1$: $T_c=0.01676$, for $\delta=3/4$, $T_c=0.2503$, for $\delta=5/4$, $T_c=0.0954$.

These values are in good agreements with the numerical values
\cite{numeric} and also coincides to the analytical values given in
\cite{powermaxwell}.

\section{Calculating the critical exponent }

We begin from equation (\ref{feqsz}) by writing it near the critical
point (CP) and in limit    $T\rightarrow T_c$. The first step is
rewriting the solution in a perturbative scheme with respect to  the
perturbation parameter $\epsilon=<O_{+}>^2$ , as it was described by
Kanno \cite{kanno}. Near the CP, the solution of the field $\phi$ is
written as
\begin{eqnarray}
&&\phi(z)=\phi_{0}+\epsilon \chi(z)\label{pertub},
\end{eqnarray}
where $\phi_0=\kappa T_c (1-z^\eta)$. Using (\ref{psi}) and
(\ref{pertub}) in first order with respect to  the $O(\epsilon)$ we
obtain the following differential equation
\begin{eqnarray}
\chi''(z)-\frac{\eta-1}{z}\chi'(z)=E(z)\label{chi}.
\end{eqnarray}
Where
\begin{eqnarray}
&&E(z)=\frac{C_{\delta} r_{+}^2z^{-\delta-2}F^{2}(z)}{ \xi
f(z)}\phi_0^2(\phi_0)'^{2(1-\delta)} ,\ \
F(z)|_{z\rightarrow0}\approx1.
\end{eqnarray}

Since $\phi(z)=\frac{A
(T_c)^{\eta+1}}{T^{\eta}}(1-z^\eta)$,$A=\frac{d-1}{4\pi}$ writing
the solution for $\phi$ in $z=0$, we have
\begin{eqnarray}
\phi(0)=\phi_{0}(0)+\epsilon \chi(0).
\end{eqnarray}
The general solution for (\ref{chi}) is given by
\begin{eqnarray}
&&\chi \left( z \right) =\sqrt {z}(c_{{1}} J_1(x) +c_{{2}}
Y_1(x))+\sqrt {z}\pi \,{
\eta}^{2-2\,\delta}{T_{{c}}}^{4-2\,\delta}{\kappa}^{4-2\,\delta}f(x),
\label{chisol}
\end{eqnarray}
with
\begin{eqnarray}
f(x)=\alpha J_1(x)\int \!{\frac { {{\rm Y_1}\left(x\right)}
\left({z}^{ \eta} -1\right )
^{2}{z}^{2\,\eta-1/2+\delta-2\,\delta\,\eta}}{{z}^{d}-z
}}{dz}&&\nonumber\\+ \beta Y_1(x)\int \!{\frac { {{\rm
J_1}\left(x\right)} \left({z}^{ \eta} -1\right )
^{2}{z}^{2\,\eta-1/2+\delta-2\,\delta\,\eta}}{{z}^{d}-z }}{dz}.
\end{eqnarray}
Here, $x=2\sqrt {(1-\eta)z},\{J_n(x),Y_n(x)\}$ are Bessel functions
of first and second kinds.

 Finding the value of $\chi(0)$ from
(\ref{chisol}), and solving it for $<O_{+}>=\sqrt{\epsilon}$, we
obtain (we take $T_c=1$)
\begin{eqnarray}
<O_{+}>\propto
T^{d-\delta-\frac{\eta}{2}}[1-T^\eta]^{\frac{1}{2}}\label{o}
\end{eqnarray}

 When the power $\delta$
decreases, the value of the $<O_{+}>$ increases. Thus we conclude
that the effect of the power $\delta$ in $d=4$ model is in the
direction of the increase of $<O_{+}>$.

However, in the five dimensions, the analysis is a little bit
different. As we observe, in $d=5$, when the power $\delta$ increases,
the value of the $<O_{+}>$ increases. Thus, we can say that the
effect of the power $\delta$ in $d=5$ model is in the direction of the
increase of $<O_{+}>$.

\section{Calculating the low temperature DC conductivity}
In this section, we compute the low-temperature DC conductivity. We
concentrate on the general space time dimension $d$, and we will try
to calculate the conductivity $\sigma$ as a function of the rescaled
frequency
$$
\hat{\omega}=\frac{\omega}{<O_{+}>^{\frac{1}{\Delta}}},\ \
\Delta=d-1.
$$
 In this limit, the behavior of the scalar field is
\begin{eqnarray}
\psi(z)=\frac{b^{d-1}}{\sqrt{2}}z^{d-1}F(z),\ \
b\equiv<O_{+}>\label{psib}.
\end{eqnarray}

 We follow the method in \cite{Siopsis}. Assuming that there exists an
external magnetic field $A(r,t)=A(r)e^{-i\omega t}$. Note that here, the applied Maxwell field  is linear. It's not related to the non linear Maxwell's field in the bulk action. In fact, the non linearity of the Maxwell field now is stored in the background metric. As we know, the field equations have some terms which involve the exponent $\delta$. This parameter denotes the non linearity, which is hidden in the structure of the background metric and through it. Moreover, it diffuses to the dynamics of the scalar field and the Abelian gauge field $U(1)$. So, to compute the conductivity the applied external magnetic field is linear, and satisfies the usual linear Maxwell field $F_{\mu\nu}^{;\mu}=0$, which is nothing just the linear wave equation.\\
  The linear wave equation
for this field reduces to
\begin{eqnarray}
-\frac{d^2 A}{d\bar{r}^2}+V(r)A=\omega^2 A\label{a}.
\end{eqnarray}
The equation (\ref{a}), is written in the Schrodinger's form and in
terms of  the new tortoise coordinate
   $\bar{r}=-\frac{1}{r_{+}}\Sigma_{n=0}^{\infty}\frac{z^{n(d-1)+1}}{n(d-1)+1}$.
The horizon $r=r_{+}$ is located at $z=1$ or $\bar{r}\rightarrow
-\infty$. Here the potential is$V=2f(r)\psi^2(r)$. The ingoing waves
in horizon behaves as a boundary condition (BC) for solving this wave equation (\ref{a}), read
as
\begin{eqnarray}
A\sim e^{-i\omega \bar{r}}\sim
(1-z)^{-\frac{i\omega}{(d-1)r_{+}}}\label{bc}
\end{eqnarray}
The electromagnetic wave equation (\ref{a}), in the coordinate $z$
reads
\begin{eqnarray}
A_{zz}+\Big(\frac{2}{z}+\frac{f'}{2f}\Big)A_{z}+\Big(-\frac{2r_{+}^2\psi(z)^2}{z^4}+\frac{\omega^2r_{+}^2}{z^4f}\Big)A=0
\end{eqnarray}
By replacing the (\ref{psib}), we obtain
\begin{eqnarray}
&&A_{zz}+p_1(z)A_{z}+Q_1(z)A=0\label{a2},\\
&&p_1(z)=\Big(\frac{2}{z}+\frac{f'}{2f}\Big),\\ &&Q_1(z)=-\frac{r_{+}^2b^{2(d-1)}z^{2(d-1)}F^2(z)}{z^4}+\frac{\omega^2r_{+}^2}{z^4f}.
\end{eqnarray}
To keep the boundary conditions, we put
$$
A=(1-z)^{-\frac{i\omega}{(d-1)r_{+}}}e^{-\frac{i\omega z}{(d-1)r_{+}}}\Theta(z)
$$
Substituting this ansatz in the (\ref{a2}), we obtain
\begin{eqnarray}
&&\Theta''+P(z)\Theta'+Q(z)\Theta=0,\label{Theta}\\&&
P(z)=\,{\frac {\,r_+ \left( 1-z \right)  \left( d-1 \right) p_{{1}}
 \left( z \right) +2i\omega\,z}{r_+ \left( 1-z \right)  \left( d-1
 \right) }}
\\&&
Q(z)={\frac {iz \left( 1-z \right) \omega\,r_+ \left( d-1 \right) p_{{1}}
 \left( z \right) +{r_+}^{2} \left( d-1 \right) ^{2} \left( 1-z
 \right) ^{2}Q_{{1}} \left( z \right) +\omega\, \left(  i\left( d-1
 \right) r_+-\omega\,{z}^{2} \right) }{{r_+}^{2} \left( d-1 \right) ^{2}
 \left( 1-z\right) ^{2}}}.
\end{eqnarray}
By imposing the regularity condition on wave function $\Theta$ at the black hole horizon $z=1$, we obtain the following auxiliary boundary condition
\begin{eqnarray}
0&=&\Theta'(1) \lim_{z\rightarrow1} \Big(\frac{1}{2}\,r_+ \left( 1-z \right)  \left( d-1 \right) p_{{1}}
 (z) +i\omega\,z\Big)+\nonumber\\&&\Theta(1) \lim_{z\rightarrow1}\Big(iz (1-z) \omega\,r_+ (d-1) p_{{1}}
 (z) +{r_+}^{2}(d-1) ^{2}(1-z) ^{2}Q_{{1}}(z) +\omega\, (  i( d-1
 ) r_+-\omega\,{z}^{2} ) \Big)
\end{eqnarray}
Explicitly, by computing the limits, we have
\begin{eqnarray}
i \left( -r_++i\omega+r_+d \right) \omega\Theta(1)=0.
\end{eqnarray}
One possibility is $\Theta(1)=0$. Another $\omega=ir_+ (d-1)$. But the last case from (\ref{bc}) leads to the
\begin{eqnarray}
A\sim e^{-i\omega \bar{r}}\sim, (1-z)\label{bc1}
\end{eqnarray}
which has no meaning as the ingoing wave toward the horizon. So we
impose $\Theta(1)=0$. Since we are working in the low-temperature
limit, we take the limit $b\rightarrow\infty$, so we rescale the
coordinate $z$ by $z\rightarrow\frac{z}{b}$. Also we must put one
suitable trial form for $F(z)$, for the case of
$\Delta=d-1>\frac{3}{2},\ \ d>2$ we put $F(\frac{z}{b})\rightarrow
F(0)=1$. We rescale (\ref{Theta}), so we obtain
\begin{eqnarray}
\Theta''+bP(z/b)\Theta'+b^2Q(z/b)\Theta=0,\ \ '=\frac{d}{d(z/b)}\label{Thetab}
\end{eqnarray}
Finally, we obtain
\begin{eqnarray}
\Theta''+\frac{z}{r_+^2}(1+\frac{2ir_{+}\hat{\omega}}{d-1})\Theta'+\frac{3i\hat{\omega}}{r_{+}(d-1)}\Theta=0.\label{beq}
\end{eqnarray}
The general solution for (\ref{beq}) reads
\begin{eqnarray}
&&\Theta(z)=z{{\rm e}^{-{\frac {{z}^{2} ( \frac{1}{2}+i\hat{\omega}\,r_+ ) }{{r_+}^{2}}}}}\Big(c_+M(\mu,\nu,\frac{1}{2}\,{\frac {( 1+2\,i\hat{\omega}\,r _+) {z}^{2}}{{r_+}^{2}}})+c_-U(\mu,\nu,{\frac {( 1+2\,i\hat{\omega}\,r _+) {z}^{2}}{{r_+}^{2}}})\Big)\label{asym}\\
&&\mu=\frac{1}{2}\,{\frac {2\,d-2+4\,i\hat{\omega}\,r_+d-7\,i\hat{\omega}\,r_+}{ ( 1+2\,i\hat{\omega}
\,r_+ )  ( d-1 ) }},\ \ \nu=\frac{3}{2}.
\end{eqnarray}
The DC conductivity in the low temperature limit is defined by
\begin{eqnarray}
\sigma(\hat{\omega})=\frac{i}{\hat{\omega}}\frac{\Theta'(-\hat{\omega}^2)}{\Theta(-\hat{\omega}^2)}
\end{eqnarray}
We need to the asymptotic limit of the (\ref{asym}). Indeed, we
guess that
$\lim_{z\rightarrow\infty}\Theta(z)\approx\Theta(-\hat{\omega}^2) $
where here the $\hat{\omega}$ is the quasinormal modes, locates on
the real axis, so we guess $\Re[{\sigma(\hat{\omega})}]=0$ except at
the poles of $\Im[{\sigma(\hat{\omega})}]$ where as
$\Re[{\sigma(\hat{\omega})}]\approx\delta(\hat{\omega})$. First, by
imposing $\Theta(1)=0$ we have
\begin{eqnarray}
&&\Theta(z)=c_-\frac{z{{\rm e}^{-{\frac {{z}^{2} ( \frac{1}{2}+i\hat{\omega}\,r_+ ) }{{r_+}^{2}}}}}}{U(\mu,\nu,{\frac { 1+2\,i\hat{\omega}\,r _+ }{{r_+}^{2}}})}\times\Big[M(\mu,\nu,{\frac {( 1+2\,i\hat{\omega}\,r _+) {z}^{2}}{{r_+}^{2}}})U(\mu,\nu,{\frac { 1+2\,i\hat{\omega}\,r _+ }{{r_+}^{2}}})\nonumber\\&&-U(\mu,\nu,{\frac {( 1+2\,i\hat{\omega}\,r _+) {z}^{2}}{{r_+}^{2}}})M(\mu,\nu,{\frac { 1+2\,i\hat{\omega}\,r _+ }{{r_+}^{2}}})\Big]
\end{eqnarray}
Here $M(\mu,\nu,x), U(\mu,\nu,x)$ denote the Kummer functions\cite{kummer}.

The quasinormal modes are the solutions of the following equation
\begin{eqnarray}
&&U(\mu,\nu,{\frac { 1+2\,i\hat{\omega}\,r _+}{{r_+}^{2}}})=M(\mu,\nu,{\frac { 1+2\,i\hat{\omega}\,r _+ }{{r_+}^{2}}})
\end{eqnarray}
Which has no closed, analytical solution and can be solved just numerically.

\section{Calculating the zero temperature  conductivity}
In this section, we calculate analytically the conductivity
$\sigma(\hat{\omega})$ in the zero-temperature $T=0$ limit. This
case corresponds to the limiting case $b\rightarrow\infty$. We begin
by writing the (\ref{a}) in the coordinate $z'=z/b$
\begin{eqnarray}
&&A_{z'z'}+\frac{1}{2z'}\Big[\frac{2-(d+1)(bz')^{d-1}}{1-(bz')^{d-1}}\Big]A+\frac{r_{+}^2}{(bz')^4}\Big[\frac{b^2z'^2\hat{\omega}^2}{1-(bz')^{d-1}}\nonumber\\&&-(b^2z')^{2(d-1)}F(bz')^2\Big]A=0\label{az'}
\end{eqnarray}
By taking the limit $b\rightarrow\infty$ we've
\begin{eqnarray}
A_{z'z'}+\frac{d+1}{2z'}A_{z'}+r_{+}^2\Big[-\frac{\hat{\omega}^2}{(bz')^{d+1}}-b^{4(d-2)}z'^{2(d-3)}\Big]A=0\label{asymaz'}
\end{eqnarray}
In (\ref{asymaz'}) we set $\lim_{b\rightarrow\infty} F(bz')=F(0)=1$. The closed form of the exact solution for (\ref{asymaz'}) depends on the value of the $d$. In the below we list the solutions for special dimensions $d=2,d=3$.
\begin{eqnarray}
&&A(z')=Cz'^{-1/4}N_{-\frac{\sqrt{1+16r_{+}^2}}{2}}(\frac{2i\hat{\omega}r_{+}}{b^{3/2}\sqrt{z'}}),\ \ d=2\label{d=2}\\
&&A(z')=\frac{H(0,\beta,\delta,\gamma,\frac{z'^2+1}{z'^2-1})}{\sqrt{z'}}\Big(c_1+c_2\int \frac{dz'}{zH(0,\beta,\delta,\gamma,\frac{z'^2+1}{z'^2-1})}\Big),\ \ d=3\label{d=3}
\end{eqnarray}
Here
\begin{eqnarray}
\beta =-\frac{1}{4}\,{\frac {4\,{b}^{8}{r_+}^{2}+{b}^{4}+4\,{r_+}^{2}{\hat{\omega}}^{2}}{{b}^{4
}}},\ \ \delta=-2\,{\frac {{r_+}^{2} \left( {b}^{8}-{\hat{\omega}}^{2} \right) }{{b}^{4}}},\ \ \gamma=-\frac{1}{4}\,{\frac {4\,{r_+}^{2}{\hat{\omega}}^{2}-{b}^{4}+4\,{b}^{8}{r_+}^{2}}{{b}^{4
}}}
\end{eqnarray}
Here $N_{\nu}(x)$ is the second kind of the Bessel function and $H(0,\beta,\delta,\gamma,x)$ denotes the  Heun doubleconfluent function \cite{heun}. By expending the (\ref{d=2},\ref{d=3}) in series in the form
$$
A(z')=A_0+A_2(bz')^2-...
$$
The conductivity can be computed via the following simple formula
\begin{eqnarray}
\sigma(\hat{\omega})=\frac{2}{i\hat{\omega}}\frac{A_2}{A_0}+\frac{i\hat{\omega}}{2}\label{sigma}
\end{eqnarray}
So, by computing the asymptotic series, the conductivity will be determined by the analytical expression.\\
It's appropriate here to present the explicit form of the
conductivity for one case. We choose the case given by (\ref{d=2}).
In the zero-temperature limit, we treat the horizon size very tiny,
so in application we take $\sqrt{1+16r_{+}^2}\approx1$. Also, it's
better we define $x=\frac{2\hat{\omega}r_{+}}{b^{3/2}\sqrt{z'}} $,
which by the definition of the $z'$ reads as $x=\frac{2\omega
r_{+}}{b\sqrt{z}}$. When $b\rightarrow\infty$, then $x<<1$, so, we
can expand the Bessel function in terms of the small argument $x$,
and hence rewrite the (\ref{d=2}) as the following form
\begin{eqnarray}
A(z)=C\sqrt[4]{\frac{b}{z}}N_{-\nu}\Big(ix\Big),\ \
\nu=\frac{1}{2},\ \ x=\frac{2\omega r_{+}}{b\sqrt{z}}\label{d=2.2}.
\end{eqnarray}
Remembering the following identities
\begin{eqnarray}
N_{-\nu}(ix)=\frac{i^{-\nu-1}I_{-\nu}(x)\cosh(\nu x)-i^{\nu-1}I_{\nu}}{\sinh(\nu x)}
\end{eqnarray}
Also, we know that
\begin{eqnarray}
I_{\frac{1}{2}}(x)=\sqrt{\frac{2}{\pi x}}\cosh(x),\ \ I_{-\frac{1}{2}}(x)=\sqrt{\frac{2}{\pi x}}\sinh(x),
\end{eqnarray}
So, we have
\begin{eqnarray}
N_{-\frac{1}{2}}(ix)=\sqrt{\frac{2}{i\pi x}}\Big[\frac{i\cosh(x)\cosh(\frac{x}{2})-\sinh(x)}{\sinh(\frac{x}{2})}\Big]
\end{eqnarray}
Also, in limit $x<<1$,we  have
$$
\frac{i\cosh(x)\cosh(\frac{x}{2})-\sinh(x)}{\sinh(\frac{x}{2})}= {\frac {2\,i}{x}}-2+\frac{7}{6}\,ix-\frac{1}{4}\,{x}^{2}+{\frac {
59}{360}}\,i{x}^{3}+O ( {x}^{4})
$$
\begin{figure}
\centering
\includegraphics[scale=0.3] {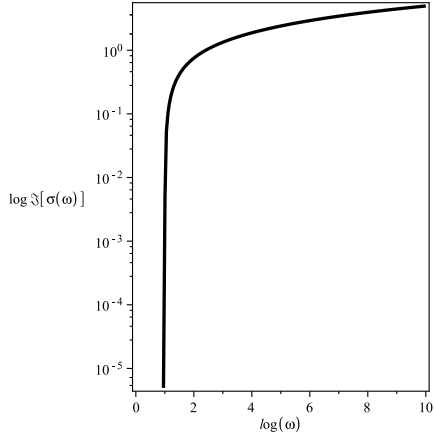}
  \caption{ Variation of the  $\log(\Im[{\sigma(\hat{\omega})}])$ for zero temperature limit  in d=2.}
  \label{1.eps}
\end{figure}

so for the electromagnetic field we've
\begin{eqnarray}
A(z)\approx C\frac{\sqrt[4]{b}}{\sqrt{i\pi r_{+}\hat{\omega}}}\Big[\frac{i\sqrt{z}}{\hat{\omega}r_{+}}-2+\frac{7}{3}i\frac{\hat{\omega}r_{+}}{\sqrt{z}}-\frac{\hat{\omega}^2r_{+}^2}{z}\Big]
\end{eqnarray}
Finally, we obtain the conductivity using (\ref{sigma}) by the following simple formula
\begin{eqnarray}
\sigma(\hat{\omega})=\frac{1}{2i\hat{\omega}}+\frac{i\hat{\omega}}{2}\label{sigma-d=2}
\end{eqnarray}
Where as we guess $\Re[{\sigma(\hat{\omega})}]=0$. The figure shows $\Im[{\sigma(\hat{\omega})}]$ as a function of the $\hat{\omega}$ for zero temperature case and in $d=2$.
The (\ref{sigma-d=2}) gives the expression for DC conductivity in the zero temperature limit in $AdS_2/CFT_1$ model. Indeed, it's comparable with the numerical results of the previous papers about one dimensional holographic superconductors \cite{Ren}.

\section{Conclusion}
In this paper, we investigated the analytical properties of a
holographic superconductor with power Maxwell's field. We studied
the problem in the probe limit. We observed that it is possible to
find the critical temperature  $T_c$ and the condensation $<O_{+}>$
and the conductivity  $\sigma (\omega)$ via Sturm-Liouville
variational approach. We concluded that in $d=4$, when the power
$\delta$ decreases, the value of the $<O_{+}>$ increases. Thus we
can say that the effect of the power $\delta$ in $d=4$ model is in
the direction of the increase of $<O_{+}>$. In $d=5$ when the power
$\delta$ increases, the value of the $<O_{+}>$ increases. Thus, we
can say that the effect of the power $\delta$ in $d=5$ model is in
the direction of the increase of $<O_{+}>$. Further, we analytically
deduced the low temperature and the zero-temperature DC conductivity
$\sigma$ as a function of the
$\hat{\omega}=\frac{\omega}{<O_{+}>^{\frac{1}{\Delta}}}$. Our work
helps to giving a better understanding of some unfamiliar effects of
the holographic superconductors both in different space dimensions
and too on the effect of the non linearity in Maxwell's strength
field.



\end{document}